\documentclass{aa}
\usepackage{natbib}
\usepackage{txfonts}
\usepackage{graphicx}  
\bibpunct{(}{)}{;}{a}{}{,}
\usepackage[english]{babel}
\usepackage{ulem}

\newcommand{\rs}{\mbox {$\rm R_{\odot}$}}
\newcommand{\ms}{\mbox {$\rm M_{\odot}$}}

\newcommand{\md}{\mbox {$\dot{M}$}}

\newcommand{\myr}{\mbox {~${\rm M_{\odot}~yr^{-1}}$}}
\newcommand{\es}{\mbox {~erg s$^{-1}$}}
\def\apgt{\ {\raise-.5ex\hbox{$\buildrel>\over\sim$}}\ }
\def\aplt{\ {\raise-.5ex\hbox{$\buildrel<\over\sim$}}\ }

\def\spose#1{\hbox to 0pt{#1\hss}}
\def\simless{\mathrel{\spose{\lower 3pt\hbox{$\mathchar"218$}}
        \raise 2.0pt\hbox{$\mathchar"13C$}}}
\def\simgreat{\mathrel{\spose{\lower 3pt\hbox{$\mathchar"218$}}
        \raise 2.0pt\hbox{$\mathchar"13E$}}}
\def\lta{\mathrel{\spose{\lower 3pt\hbox{$\mathchar"218$}}
        \raise 2.0pt\hbox{$\mathchar"13C$}}}
\def\gta{\mathrel{\spose{\lower 3pt\hbox{$\mathchar"218$}}
        \raise 2.0pt\hbox{$\mathchar"13E$}}}
\def\msol{\rm M_\odot}

\begin{document}
\title{The origin and fate of short-period low-mass black-hole binaries}

\author{L. R. Yungelson\inst{1,2,3,7}
          \and
           J.-P. Lasota\inst{3}
            \and
            G. Nelemans\inst{4,5}
             \and
              G. Dubus\inst{3,6}
               \and
               E.P.J.~van~den~Heuvel\inst{7}
                \and
                J. Dewi\inst{5,7}
                 \and
                 S. Portegies Zwart\inst{7,8}
}

\offprints{J.-P. Lasota, lasota@iap.fr}

\institute{Institute of Astronomy of the Russian Academy of
            Sciences, 48 Pyatniskaya Str., 119017 Moscow, Russia
            \and
            Isaac Newton Institute, Moscow branch, 13 Universitetskii pr., Moscow, Russia
             \and
               Institut d'Astrophysique de Paris, UMR 7095 CNRS,
Universit\'e Pierre et Marie Curie, 98bis Bd Arago, 75014 Paris,
France
                \and
         Department of Astrophysics, Radboud University Nijmegen, Toernooiveld 1, NL-6525 ED Nijmegen, the Netherlands
                 \and
                  Institute of Astronomy, University of Cambridge, Madingley Road, CB3 0HA, Cambridge, UK
                   \and
                      Laboratoire Leprince-Ringuet, UMR 7638 CNRS, Ecole Polytechnique, 91128 Palaiseau, France
                    \and
                    Astronomical Institute ``Anton Pannekoek'',
                      Kruislaan 403, NL-1098 SJ Amsterdam, the Netherlands
              \and
              Section Computational Science, Informatics Institute,
              Kruislaan 403, NL-1098, SJ Amsterdam, the Netherlands
              }

\date{received \today}
\titlerunning{Low-mass binaries and SXT}
\authorrunning{L.R. Yungelson et al.}

\abstract{We present results of a population synthesis study for
semidetached short orbital period binaries which contain low-mass
($\aplt1.5\,\ms$) donors and massive ($\gtrsim 4\,\ms$) compact
accretors, which in our model represent black holes. Evolution of
these binaries is determined by nuclear evolution of the donors and/or
orbital angular momentum loss due to magnetic braking by the stellar
wind of the donors and gravitational wave radiation.  According to our
model the estimated total number of this type of black-hole binaries
in the Galaxy is $\apgt 10000$.  If the magnetic braking removing
angular momentum in semidetached black-hole binaries is described by
the Verbunt \& Zwaan formula, the model predicts that among them
around 3000 systems with periods $>$ 2 hours would be transient. In
addition one finds several hundreds of luminous and stable systems
with periods between 3 and 8 hours.  Several dozens of these bright
systems should be observed above the \textit{RXTE ASM} sensitivity
limit. The absence of such systems implies that upon Roche-lobe
overflow by the low-mass donor angular momentum losses are reduced by
a factor more than 2 with respect to the Verbunt \& Zwaan
prescription. In such a case short period black-hole soft X-ray
transients (SXT) may have donors that overflow the Roche lobe in the
core hydrogen-burning stage. We show that it is unlikely that the
transient behaviour of black-hole short-period X-ray binaries is
explained by the evolved nature of the stellar companion. On the other
hand a substantial fraction of black-hole binaries with periods $>$ 3
hours could be faint with truncated, stable cold accretion discs as
proposed by Menou et al. Most of the semidetached black-hole
binaries are expected to have periods shorter than $\sim$ 2
hours. Properties of such, still to be observed, very small mass-ratio
$(q=M_2/M_1 < 0.02)$ binaries are different from those of their longer
period cousins.

\keywords{binaries: close -- binaries: evolution -- X-rays: binaries}
}

\maketitle

\section{Introduction}
\label{sec:intro}

All known low-mass X-ray binaries containing black hole components
(massive compact accretors, hereafter LMBHB) are transient
\citep{mclrbh}. The instability responsible for this, often
intermittent, behaviour is apparently the same that drives dwarf-nova
outbursts \citep[see][ for a detailed review of the ``disc instability
model'' -- DIM]{lasota01} but in the case of LMBHB irradiation of the
outer disc plays a crucial role
\citep{vanpar96,dubhamlas99,dubhamlas01}. For this instability to
operate, rather low mass-transfer rates are required ($\aplt 10^{-10}
- 10^{-9}$\,\myr). This conjecture has led to the suggestion that
all LMBHB are formed via a peculiar evolutionary path, in which the
donor stars start mass transfer to the black hole only at the very end
of their main-sequence lifetime
\citep[e.g.][]{kkb96,ef98,ergmasarna01}. It is therefore of interest
to check which models of binary evolution are capable
of reproducing the transient nature of the LMBHB population and their
observed (or deduced from observations) mass-transfer rates. In
addition population synthesis models can be used to test the
statistics of the model versus the observations and the existence
of a hidden (quasi)stable population of LMBHB postulated by
\citet{mnl99}.

Among the 17 confirmed black hole binaries, (i.e. with a
measured mass-function larger than $\sim 3$ \ms) ten are
short-period ($P_{\rm orb}\aplt 1$\,day) systems
\citep{orosz16,mclrbh}. Their evolution should be similar to that of
cataclysmic variables (CV) which prompts to apply to LMBHB the model
exploited for CV: evolution driven mainly by angular momentum loss
via magnetically coupled stellar wind (MSW) and gravitational waves
radiation (GWR). This is a non-trivial exercise because of the
well-known problems of the ``standard model" of CV evolution, e.g.
the difficulty in reproducing the period gap and period minimum
observed in the distribution of CVs \citep[see e.g.][]{kolb_goet}.

The criteria for the instability in the DIM depend mainly on the
mass transfer rate, which is set by the period of the system and
mass of the accreting body, i. e., along its evolutionary track a
binary may move trough stable and unstable states.

Therefore the evolution of LMBHBs provides an interesting
challenge to our understanding of both the stellar evolution and the
accretion-disc physics. This prompted us to carry out a population
synthesis study for low-mass binaries with massive companions based
on a systematic investigation of the evolution of the latter systems
under different assumptions on the angular momentum loss from the
system. As the next step we attempted to infer the behaviour of  a
constructed population with respect to the stability criteria in DIM
and to verify, whether it is possible that the Galaxy harbours a
population of LMBHB with faint cold and stable discs.  Population
synthesis results are presented in Section \ref{sec:popsynth} and
compared to observations in Section \ref{sec:transient}. Discussion
and Conclusions follow.

\section{Formation of low-mass black-hole binaries and their current
  Galactic population}
\label{sec:popsynth}

\subsection{Formation of binaries containing a black hole and a
    low-mass star}

The formation of close binaries containing a black hole and a low-mass
  star is is still not well understood. The progenitor of the black
  hole must have been a very massive star, with heavy mass loss,
  which went through a supernova explosion. It is remarkable that
  apparently a low-mass companion can survive such violent
  evolution. The ``standard'' formation channel for such binaries is
  an extension of the original formation scenario proposed for
  low-mass X-ray binaries with a neutron star accretor \citep{heu83}:
  a massive star with a low-mass companion evolves and the low-mass
  star spirals into the giant in a common-envelope phase. The product
  of common-envelope stage is a compact binary consisting of the
  helium core of the massive star (a Wolf-Rayet star) and the low-mass
  companion.  Neither the small amount of matter lost in supernova
  explosion in which the neutron star is formed nor a natal kick
  imparted to neutron star disrupt the system, so the binary survives
  and, due to angular momentum losses, becomes tighter and tighter
  until mass transfer ensues. In order to end with a black hole, one
  simply needs a more massive star to start with
  \citep[e.g.][]{macclintock86,khp87}. Unfortunately, the above
  scenario contains many very uncertain aspects: the amount of mass
  loss of massive stars, the outcome of the common-envelope phase, the
  amount of mass loss of the helium core after the common-envelope
  phase and the mass loss and possible asymmetric kick induced at the
  supernova explosion \citep[see
  e.g.][]{lan89b,ity95b,wl99,kal99,fryer_kal_bh01,nh01, justham_bh06}.

\subsection{Evolution of black-hole low-mass X-ray binaries: Evolutionary computations}

In order to be able later to compare the properties of the observed
  black-hole low-mass X-ray binaries with our model, we have made
  detailed calculations of a grid of evolutionary sequences for
  different black hole and companion star masses and different initial
  periods.   Depending on these parameters, the low-mass companions
  can overflow (if ever) their critical lobes at different
  evolutionary stages. Consideration of a range of initial periods is
  important, in particular to assess suggestion than transient
  low-mass black-hole X-ray binaries must all harbour donor-stars that
  almost exhausted hydrogen in their cores, e.g.
  \citep{kkb96,ef98,ergmasarna01}.

      Evolutionary computations were carried out by means of  the TWIN version
 of \citet[][priv. comm.  2003]{egg71} evolutionary
  code. The most important updates of the code concerning equation of
  state, nuclear reaction network and opacity are described by
  \citet{pteh95}. Our version of the code implements angular momentum
  loss (AML) via a magnetically coupled stellar wind (MSW), following
  the model suggested by \citet{vz81} and via gravitational wave
  radiation \citep[GWR,][]{ll71}. We should mention specifically that,
  at difference with other codes which include AML via MSW, it is
  assumed that a critical minimum depth of the convective zone
  $Z_{\mathrm{conv}} > 0.065\,\rs$ is required for sustaining a
  significant surface magnetic field. Within assumptions used in the
  code, this limits the masses of main-sequence stars which may have
  MSW to lower than 1.6\,\ms. On the other hand, since it is
  \textit{believed} that the magnetic field is anchored in the radiative
  core, MSW is switched-off as soon as convective envelope penetrates
  to $0.2\,R_{\mathrm{star}}$ (``disrupted'' magnetic braking
  model). For initially unevolved MS-stars MSW terminates when their
  mass decreases to $\sim 0.3$\,\ms.

The \citet{vz81} AML law can be written as
\begin{equation}
\frac{dJ}{dt} =-0.5 10^{-28}f^{-2} k^2 M_d R_d^4 \omega^3\rm\,dyn\,
cm,
\label{eq:vz}
\end{equation}
where $k^2\sim 0.1$ is the gyration radius of the donor star and
$M_d$, $R_d$ and $\omega$ are respectively its mass, radius and
rotation angular frequency, $f$ is a parameter calibrating the strength of MSW AML.
In the evolutionary code, the gyration
radius is estimated for every model.

Solar chemical composition of the models was assumed. Trial
computations for donors with metallicity $ Z$=0.008 did not reveal any
serious dependence of results on $Z$.  Different metallicities might
also influence the mass of the black hole, as the mass loss rates of
massive stars are expected to scale with $Z^{0.5}$. However, what
really matters for our purpose is the pre-SN mass. Since both mass
loss rates for MS- and WR-stars are highly uncertain we neglect the
Z-dependence of stellar winds.  Comparison of our pre-SN masses with
the results of trial computations with the \citet{hurley2000} code
which employs a different combination of stellar winds shows that
pre-SN masses are the same within a few solar masses for metallicities
of 0.02 and 0.008. We neglected the possible influence of an
irradiation induced stellar wind \citep{tl93} on the donor evolution.
This is justified since irradiation feedback is important only for
systems with $M \sim \ms$ donors \citep{bun_rit05} which comprise a
negligible minority of population under consideration.

In codes describing CV evolution the AML is usually normalized in a
way to obtain the correct location of the period gap \citep[see
e.g.][]{kolb_strasb}. Since no gap has been detected for the LMXB
orbital distribution we left our code uncalibrated. Let us add that at
variance with cataclysmic variables, in systems with massive primaries
the donors are only slightly out of thermal equilibrium and the width
of the calculated ``period gap'' is small, compared to CVs. For the
models that overflow Roche lobes when they have already burnt a
significant fraction of hydrogen in their cores the gap disappears
(see Fig. \ref{fig:example} below).

The grids of evolutionary models were computed for initial secondary
masses ($M_{20}$) of  $0.4 - 1.6$\,\ms\ and initial black hole
masses ($M_{10}$) of 4, 7, and 12\,\ms. In the first set of
computations a ``standard'' model of evolution of low-mass binaries
with AML via MSW ($f=1$ and disrupted magnetic braking) and GWR was
assumed (henceforth, model A). Since population synthesis under
these ``standard'' assumptions did not provide satisfactory results
(see below), in the second set of computations we assumed {\it ad
hoc} that MSW does not operate in semidetached systems with
black-hole components (henceforth, model B). In
model B only GWR operates after RLOF. We also used for comparison a model
similar to model A in which the MSW angular-momentum loss rate
operated permanently, but reduced by a factor $\sim 2$ ($f=1.37$
-- model C). Evolution models A, B, and C give rise to 13737,
12089, and 11659 short-period semidetached systems, respectively.
Corresponding  LMBHB-population models are compared in  detail in
Sec.~\ref{sec:transient}.

The models for low-mass donors typically ceased to converge at
$M_2^{min} \simeq 0.03$\,\ms\ in evolutionary tracks for models A and
C and at $M_2^{min} \simeq 0.06$\,\ms\ for model B. This happens
because of the absence in the code of adequate tables of EOS and
opacities for that low-mass stars the slight difference in the limiting
masses stems from some differences in the versions of the code used
for models A and B, C; in our results this only slightly influenced
$M_2^{min}$ and position of orbital period minimum). For population
modeling, all tracks were truncated at $M_2^{min}$=0.02\,\ms\ or
0.06\,\ms, respectively. However, already at $M_2$=0.06\,\ms\ the mass
ratio of components $q < 0.02$, the circularization radius of
accretion stream for such low $q$ becomes greater than the outer
radius of the accretion disc and it is unclear, whether mass transfer
happens at all (see \ref{sec:low_q} for more details).
 Thus, although evolutionary computations were continued beyond $q
\approx 0.02$ with the same formalism as for larger $q$, the results may
be not applicable to any physical systems.

\subsection{The Galactic population of low-mass binaries with black-hole companions}

\begin{figure}
\center
\includegraphics[angle=-90,width=\columnwidth]{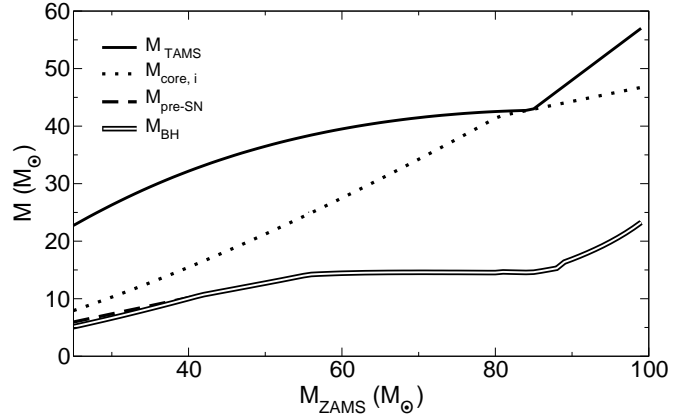}
\caption[]{Initial mass -- final mass relations for massive stars in
binaries. The lines show the mass of stars at the end of the main
sequence (TAMS), the initial  mass of helium core ${\rm M}_{\rm
core, i}$, the pre-supernova mass ${\rm M}_{\rm pre-SN}$, and the
final black-hole mass ${\rm M}_{\rm BH}$.}
\label{fig:IMFM}
\end{figure}

We modeled the Galactic population of black-hole binaries using the
population synthesis code \textsf{SeBa} \citep{pv96} with updates
described in \citet{py98,nyp+01,nyp04}. The important points to note
are that we assume a 50\% binarity rate (2/3 of stars in binaries), an
IMF after \citet{ktg03}, an initial distribution of separations in
binary systems ($a$) flat in $\log a$ between contact and 10$^6$
R$_\odot$, a flat mass ratio distribution, and initial distribution of
eccentricities of orbits $\Xi(e)=2e$.

A summary of the assumptions about the evolution of massive stars
and Wolf-Rayet stars that we use in our code is
 shown in Fig.~\ref{fig:IMFM} where, for components of close
binaries with initial masses of $M_0=25 - 100$\,\ms, we show their
masses at the end of the main sequence (TAMS), their initial helium
core mass, their pre-supernova mass and the final black-hole
mass\footnote{ The overwhelming majority of close binaries donor
stars lose their hydrogen envelope through mass transfer after the
main sequence, but before the onset of helium core burning.  For
such binaries the final masses are rather insensitive to the initial
period of the binary  and the initial $q$.}. We assume
that the stellar wind mass-loss  rate increases in time and that in
their total lifetime stars may lose an amount of matter that
increases with initial mass: $0.01 M_{\mathrm i}^2$. If the star
loses its hydrogen envelope, we switch to the Wolf-Rayet stellar mass-loss
prescription. Mass loss on the main-sequence causes stars with
masses of about 85\ms\ to lose all their hydrogen before they reach
the TAMS. For more massive stars the mass loss is so uncertain that
we assume for simplicity that they lose  42\,\ms\ on the main
sequence, leading to the peculiar upturn of the solid line in
Fig.~\ref{fig:IMFM}.  However, these extremely massive stars are so
rare that they
do not contribute to the population of low-mass black-hole binaries.
 For the
common-envelope phase we use the standard prescription
\citep{web84,khp87}, with the efficiency and structure parameters
$\alpha$ and $\lambda$ combined: $\alpha \lambda = 2$. For the mass
loss by Wolf-Rayet stars formed after the common-envelope phase, we
use the law derived by \citet{nh01}, which is based on the compilation
of estimated mass-loss rates of observed Wolf-Rayet stars of
\citet{nl00}.

The formation of black holes in the code follows the fall-back
prescription of \citet{fryer_kal_bh01} whose basic assumption is
that a constant fraction of the supernova explosion energy is used
to unbind the envelope of the star.  At difference to
\citet{fryer_kal_bh01}, who consider the explosion energy as a
function of pre-supernova mass, we keep it fixed at $10^{50}$ ergs,
which is within the expected range but favours formation of rather
massive black holes (up to $\sim15$ \ms).  For systems that are
relevant for this study, i.e. naked helium cores, the prescription
we use actually almost invariably results in complete collapse of
the pre-supernova object: the CO core is very tightly bound while
around this core only a small layer of helium is left because of the
strong stellar wind during the preceding Wolf-Rayet phase. We assume
that black holes receive a small asymmetric kick at formation
\citep[e.g.][]{jn04,whl+04,gcp+04}, which is taken from a
\citet{pac90} and \citet{har97} velocity distribution with a
dispersion of 300 km~s$^{-1}$, but scaled down with the ratio of the
black hole mass to a neutron star mass.  The population of
short-period low-mass black-hole binaries is not sensitive to the 
assumed kick distribution, since scaled down with $M_{\rm bh}/M_{\rm
ns}$-ratio kick velocities are too small to disrupt close binaries
in SN explosions. This was confirmed by a test run assuming
Maxwellian kick distribution for pulsars with $\sigma=265$\,km/s
\citep{hobbs2005}. The total number of systems that filled Roche lobe in
Hubble time and evolved to shorter periods (in models A and B, see
below) changed by several dozens only.

The model of the Galactic population of detached low-mass binaries
with black-hole companions is then produced by folding the results of
approximate evolutionary computations for a set of 250000 initial
binaries with $M_{10} \geq 25$\,\ms\ with a time and position
dependent model for the star formation history of the Galaxy that
is based on results of \citet{bp99} and a model for the distribution of stars in
the Galaxy \citep[see][]{nyp04}. Some details of this model are
described in Appendix~\ref{appendix_model}.

With our assumptions we find that, within a Hubble time (13.5
Gyr), $\sim 49\,000$ binaries that have orbital periods below 2.0 day
and contain black holes accompanied by main-sequence stars less
massive than 1.6\,\ms were formed in the Galaxy.\footnote{This number
is $\sim$0.01\% of all binaries with black-hole components formed in
Hubble time.}  We restrict ourselves to $M_{20} \leq 1.6$\,\ms\, since
for more massive stars magnetic braking is not supposed to
operate. Out of these $\sim 49\,000$ binaries, $\sim 17,000$ were
brought into contact in a Hubble time, but only $\sim 14\,000$ evolved
to shorter periods (Table \ref{tab:numbers}). If only AML via GWR acts
in systems with more massive donors, they experience case AB
mass-transfer\footnote{According to common classification, in ``case AB'' mass-transfer starts when the donor is still in the core-hydrogen burning stage and continues in the hydrogen-shell burning stage.}, evolve into subgiants with increase of orbital periods
and end their lives as helium white dwarfs.

\begin{figure}
\center
\includegraphics[angle=-90,width=\columnwidth]{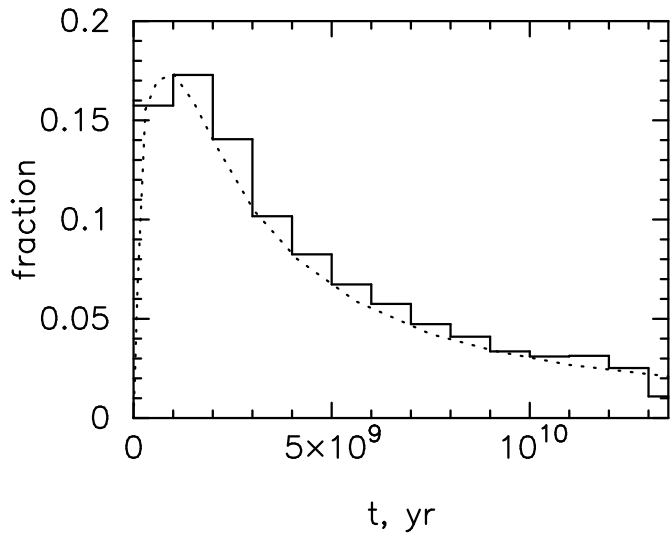}
\caption[]{Solid histogram - fractions of low-mass bh+ms binaries formed at different
epochs in the lifetime of the Galaxy. Dotted curve shows the shape of star formation history adopted in the population synthesis code. At maximum, the star formation rate (SFR) is $\simeq 15.6$\,\myr. Current SFR is $\simeq 1.9$\,\myr.}
 \label{fig:history}
\end{figure}

 Because black holes form in the first $\sim$3 Myr after their
progenitor formation, the formation history of bh+ms binaries
(Fig.~\ref{fig:history}) strictly follows the star formation history.

\begin{figure}
\center
\includegraphics[width=\columnwidth]{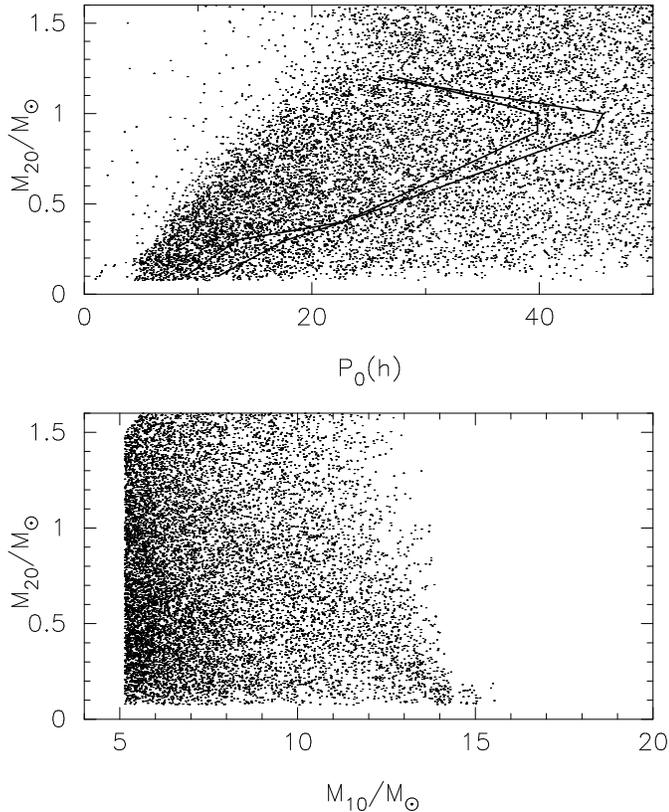}
\caption[]{Upper panel: relation between orbital periods and masses
of main-sequence companions of black holes after circularization of
the orbits.  Systems with main-sequence components that may fill
their Roche lobes in their main-sequence lifetime or in the Galactic
disc lifetime (13.5 Gyr) are located to the left of the solid
curves. Left and right solid curves correspond to $M_{\rm
BH}=4\,\ms$ and $12\,\ms$. Lower panel: relation between initial
masses of components in bh+ms systems.} \label{fig:p0m20}
\end{figure}

The upper panel of Fig. \ref{fig:p0m20} shows the relation between
initial masses of MS-components and orbital periods after
circularization of the orbits after black hole birth
events\footnote{Because of specifics of time-realizations in the
population synthesis code every dot in the plot actually represents
four or five systems with similar combination of masses and orbital
period, but born at different time. Since we present one random
realization of the model for a population of
black-hole+main-sequence star systems, all numbers given are subject
to Poisson noise.}. The lower panel represents relation between
masses of components in these systems.

\subsection{The current Galactic population of black-hole low-mass
    X-ray binaries}
\label{sec:pop_synth}

We have simulated the current Galactic population of semidetached
low-mass black-hole binaries by convolving the above-described
``underlying'' population of binaries born at different epochs 
in the
history of the Galaxy with the grids of pre-computed evolutionary
tracks for low-mass components in binaries with different combinations
of components and post-circularization (initial) orbital periods.

\begin{figure}
\center
 \includegraphics[angle=-90,width=\columnwidth]{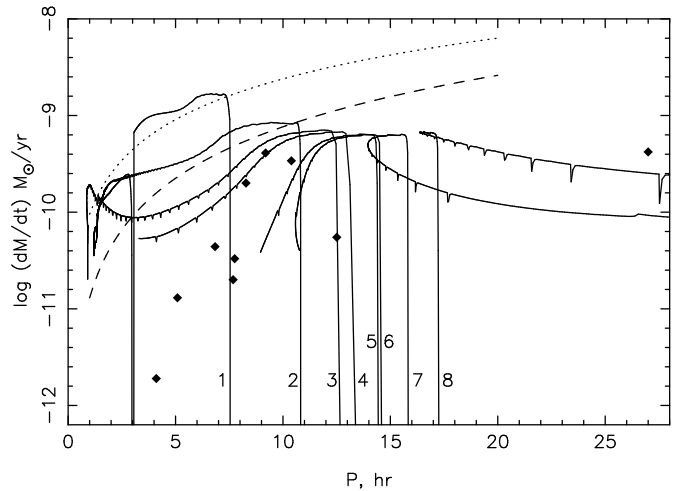}
  \caption[]{ Mass transfer rates \textit{vs.} orbital period for
systems with initial masses of components 1\,\ms\ and 7\,\ms\ that
have different periods $P_0$ after circularization of orbits. Solid
lines show tracks of donors evolving under ``standard'' assumption
that AML via MSW operates until secondaries become convective
($f$=1).  The numbers along the tracks correspond to the numbers in Table~\ref{tab:example}. The leftmost line is for $P_0=9.6$\,hr; in this system
secondary fills its Roche almost unevolved at $P \approx
7.44$\,hr. Note the presence of a tiny period gap for this system.
The systems to the right have $P_0$ from 38.4 to 45.6 hr
(respectively, at RLOF they have, either $X_c \leq 7\cdot10^{-4}$ or
helium core with $M_{\rm He} < 0.01M_\odot$, see Table
\ref{tab:example}). The dotted line shows mass transfer rate critical for
disc instability (Eq.  \ref{eq:mcrit}). The dashed line marks the
approximate upper limit to the average mass transfer rates for
short-period black-hole binaries, as given by
Eq.~(\ref{eq:mdot_adaf2}). The diamonds show estimates of
mass-transfer rates based on assumed recurrence period, see Table \ref{tab:obs}. }
 \label{fig:example}
\end{figure}

Since the features of the model population may be understood as a
consequence of different patterns of evolution of low-mass donors
(Fig. \ref{fig:example}) we will first present examples of
evolutionary tracks. The parameters of the tracks shown in
Fig.~\ref{fig:example} are listed in Table \ref{tab:example}. For
every combination of $M_{20}$ and $M_{10}$ there are systems that,
upon Roche lobe overflow (RLOF), evolve to shorter periods having
mass-transfer rates $\dot{M} \approx (10^{-9} - 10^{-10})\,\myr$; for
$M_{\rm 20} \simeq 1\,M_\odot$ this happens for systems with  an abundance of hydrogen in their center $X_c\apgt 0.05$ at RLOF.
These systems may have post-circularization
periods up to $\sim 40$\,hr (Fig.  \ref{fig:p0m20}) and they overflow
Roche lobes at periods $\aplt 12$ hr (see Table \ref{tab:example} and
Fig. \ref{fig:example}) since the period of semidetached systems
depends only on the mass and radius of the donor and the latter
changes only slightly in the hydrogen burning stage or practically
does not change in Hubble time for $M_{20} \aplt0.95\,\ms$. Systems
with hydrogen-exhausted cores (two rightmost tracks in
Fig.~\ref{fig:example},
No.~7 \& 8
 in Table \ref{tab:example}) evolve
to longer periods. This is the well known effect of ``bifurcation
period'' \citep{tfey85,pylsav89}. The bifurcation period $P_b$ defines
the right border of the position of new-born systems that evolve to
shorter orbital periods upon RLOF, as shown in Fig.~\ref{fig:example}.

\begin{table}[t!]
\caption[]{Parameters of tracks shown in Fig.~\ref{fig:example}. The
columns list initial orbital period of the system, period at RLOF,
central hydrogen abundance at RLOF, mass of helium core at RLOF.  }
\begin{center}
\begin{tabular}{rcccc}
\hline\hline
 No. & $P_0,$ & $P_c,$ & $X_c$ & $M_{\rm He0},$\\
& hour & hour & & $M_\odot$\\
\hline
 1 & 9.60  & 7.44 & 0.697  & -  \\
 2 & 38.40 & 10.80 & 0.049  & - \\
 3 & 40.80 & 12.48 & $7 \cdot 10^{-4}$ & - \\
 4 & 41.52 & 12.96 & $2 \cdot 10^{-4}$ & - \\
 5 & 42.96 & 14.40 & $2 \cdot 10^{-8}$ & - \\
 6 & 43.20 & 14.40 & $2 \cdot 10^{-9}$ & - \\
 7 & 44.40 & 16.08 & -                 & $4 \cdot 10^{-3}$ \\
 8 & 45.60 & 17.28 & -                 & $1 \cdot 10^{-2}$ \\
\hline
\end{tabular}
\end{center}
\label{tab:example}
\end{table}

In systems that are initially driven by both MSW and GWR, \md\ drops
rapidly from $\md \sim 10^{-9}$\,\myr\ to $\sim 10^{-10}$\,\myr\ after
termination of MSW.  Like in cataclysmic variables, the evolution
continues with decreasing \md\ and there exists a minimum orbital
period $\sim (70 - 80)$ min. LMBHB close to the period minimum have
yet to be observed but our population synthesis model predicts that
they might form the \textsl{majority} of LMBHB. These systems have
very low mass ratios $q<0.02$ and it is far from obvious that mass
transfer occurs in them in a similar fashion to standard compact
semidetached binaries (see Section \ref{sec:low_q}).

For the most massive stars that still have MSW and fill their Roche
lobe close to $P_b$, nuclear evolution continues to play a
certain role
after RLOF. As a result, they first evolve to shorter periods, but,
when hydrogen is almost completely exhausted in their cores, turn to
longer periods (tracks with ``loops'' in Fig.~\ref{fig:example}). In
the extreme case of $M_{20}=1.6$\,\ms\ donor, this ``u--turn'' may
happen at a period as short as $\sim 4$ hr, but such systems with
periods from a very narrow interval of initial $P_{orb}$ are extremely
rare in the parental population.

\subsubsection{Evolved companions ?}

The systems that have at RLOF~$0.05 \apgt X_c \apgt 0$ evolve to short
periods and may have from the very beginning of the Roche-lobe
overflow $\md < 10^{-9}\,\myr$. In the period range 4 -- 12 hr they
have rather low mass-transfer rates: $\md \sim 10^{-10} -
10^{-11}\,\myr$\ \citep{tfey85}. They are represented by the tracks
No. 3 -- 6 (with initial periods 40.8 -- 43.2 hr and periods at
contact 12.5 -- 14.8 hr) in the centre of Fig. \ref{fig:example}. In
the model population, they do not evolve to periods below about 4 hr,
since they have to be already significantly evolved at the instant of
RLOF which leaves them only a few Gyr of evolutionary lifetime in
semidetached state.  Since the interval of periods of these systems
and their mass-transfer rate $\dot{M}$ overlap with the those of SXT
it is tempting to assume that they provide the explanation of the
properties of these systems. However, this cannot be the case because
in the majority of systems the donors fill their Roche lobes almost
unevolved: for the systems shown in Fig. \ref{fig:example}, the ratio
of initial period intervals of systems that fill Roche lobes
``evolved'' and ``unevolved'' is $\sim 8$\%. The actual ratio of the
numbers of stars in two groups is much lower, since only systems with
$M_{20} \apgt 0.95\,\ms$\ may exhaust hydrogen in their centres in
Hubble time. This is confirmed by the upper panel of Fig.
\ref{fig:popul_las} (see below), which shows that even the upper
estimates of the average mass-transfer rates in observed SXT hardly
may be explained by the standard AML via MSW and GR model, because of
low number of systems with desired initial parameters in the parental
population and their short lifetime in the observed period range.
Simply removing all systems that will fall in the ``unevolved'' class
by fine-tuning the initial period distribution of bh+ms is difficult:
the most obvious way would be to change the common-envelope
efficiency. However we tried this by repeating our calculations with
$\alpha \lambda = 0.5$ and 5 (the standard value in our model is 2),
but the ``unevolved'' class still dominates. The number of black hole
+ low-mass MS-star systems formed in a Hubble time was, compared to
``standard'' model, 50\% lower and 15\% higher respectively for these
two calculations.\footnote{Respectively 24,000 and 58,000 systems.}

This shows that the transient nature of the short period LMBHB most likely
cannot be explained by the presence of strongly evolved
main-sequence companions as proposed by e.g.
\citet{kkb96,ef98,ergmasarna01}.

\section{Accretion in the modeled LMBHB population and its observational consequences}
\label{sec:transient}

\subsection{Bright, persistent systems}
\label{sec:dimstable1}

\subsubsection{Disc stability}

\begin{figure}
\center
   \includegraphics[angle=-90,width=\columnwidth]{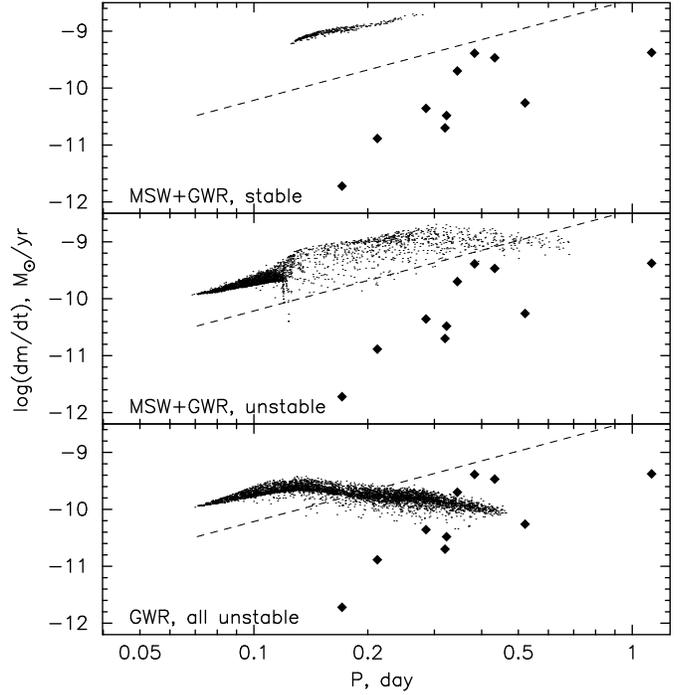}
    \caption[]{Mass transfer rates vs. orbital period for the model
population of semidetached low-mass binaries with assumed black hole
accretors. Upper panel shows systems that have irradiated discs that
are stable according to criterion given by Eq. (\ref{eq:mcrit}) in the
model for a ``standard'' assumption on AML via MSW always operating in
stars with convective envelopes ($f=1$, Model A). Middle panel shows
population of systems with unstable irradiated disks in the same
model. Lower panel presents population for the model in which MSW does
not operate in semidetached systems with black hole accretors (model
B). In this model all binaries have unstable irradiated disks.  The
break-down over number of stable and unstable systems in particular
models is listed in Table \ref{tab:numbers}.  The dashed line marks
the approximate upper limit to the average mass-transfer rates for
short-period black-hole binaries, as given by
Eq.~(\ref{eq:mdot_adaf2}), while diamonds show estimates of rates
based on assumed recurrence period of 30 years, except for V616 Mon
and IL~Lup whose recurrence times are known to be $\sim 50$ and 10
years, respectively (see Table \ref{tab:obs} and text for details).  }
\label{fig:popul_las}
\end{figure}

Figure \ref{fig:popul_las} (upper and middle panels) shows the modeled
population of semidetached low-mass binaries with assumed black hole
accretors in ``standard'' model (A) with disrupted magnetic
braking. In Figure \ref{fig:popul_las}, only systems for which DIM is
valid and stability analysis may be applied are plotted ($\sim 3000$
systems).  The remaining systems (Table \ref{tab:numbers}) have
low-mass ratios of components ($q<0.02$) and low-mass ($M_2
<0.03$\,\ms) donors. If mass transfer occurs and can be described by
the ``standard'' model of evolution driven by AML by GWR, these
systems should be concentrated around a ``period minimum'' at 70 -- 80
min and beyond it and at mass-transfer rates $10^{-10} -
10^{-11}$\,\myr.  Model A contains $\sim$ 13000 systems (Table
\ref{tab:numbers}) of which some may be luminous and non-transient
(upper panel of Fig.~\ref{fig:popul_las}). Such steady systems should
satisfy the stability criterion according to which irradiated discs
are stable if their accretion rate exceeds certain critical value
$\md_{\rm crit}^+$\ \citep{vanpar96}. According to \citet{dubhamlas99}
the latter is
\begin{equation}
\label{eq:mcrit} \md^+_{\rm crit} \approx 2.4 \times 10^{-11} M_{\rm
BH}^{-0.4} \left(\frac{R_d}{10^{10}\mathrm{~cm}}\right)^{2.1} \left(
\frac {\mathcal{C}}{5 \times 10^{-4}}\right)^{-0.5}  \myr,
\end{equation}
where $M_{\rm BH}$ is the mass of accretor in \ms, $R_d$ is the disc
outer radius, and $\mathcal{C}$ is a measure of heating of the disc
by X-rays \citep{shaksun73}; for a given disc radius $R$ the
irradiation temperature $T_{\rm irr}$ is
\begin{equation}
\label{eq:tirr} T_{\rm irr}^4=\mathcal{C} \frac{\md c^2} {4 \pi
\sigma R^2},
\end{equation}
where $\sigma$ is the Stefan-Boltzmann constant. We use
$\mathcal{C}=5 \times 10^{-4}$, following \citet{dubhamlas99} who
have found that this value is consistent with properties of
persistent low-mass X-ray sources. One should, however, keep in mind
that $\mathcal{C}$ does not have to be a constant and might e.g.
vary in time \citep{elh}.

The disc outer radius $R_d$, which depends only on the mass ratio
once scaled to the orbital separation, is estimated using Tab.~2 of
\citet{Paczynski77} who computed the largest possible stable,
periodic non-intersecting orbits. This estimate is very close to the
one obtained by \citet{pappringle77} by studying the linear response
to tidal forcing. For mass ratios $q\approx 0.02$, the disc outer
radius becomes comparable to the circularization radius $R_{\rm
circ}$ and  the $R_d$ estimate is not valid. It is then uncertain how
accretion proceeds. In what follows we concentrate on systems with
mass ratios $q \ge 0.02$ (3272 systems in model A), corresponding to
periods longer than $\ga 2$ hours, and come back to the systems with
low mass ratios in Section \ref{sec:low_q}.

\begin{table}[t!]
\caption[]{Break-down of black-hole candidate populations into
sub-groups according to accretion disc stability. Model A --
``standard'' evolution model with disrupted magnetic braking
($f=1$), Model B - model without MSW in semidetached systems. Model
C corresponds to AML by MSW reduced by a factor 2 ($f=1.37$). The
total number of systems in contact is indicated for each model. The
population is then separated into systems with $q< 0.02$ for which
description in all evolutionary calculations may be inadequate (see
 Section \ref{sec:low_q}) and systems with $q \geq 0.02$. For the
latter, the systems are further broken down into bright persistent
systems (DIM stable, Section \ref{sec:dimstable1}) and transient
systems. Some of the systems marked a ``transient" could be
cold, stable, truncated-disc systems (MNL stable, Section
\ref{sec:mnlpar}). 
\begin{center}
\begin{tabular}{lrll}
\noalign{\smallskip}
\hline\hline
  & Model A & Model B & Model C\\
\hline
Total       & 13737 & 12089 & 11659\\
$q\ge0.02$  & 3272  & 5074  & 2991\\
\hline
Hot stable  & 367  &  0     &  107\\
Transient   & 2905 &  5074  &  2884\\
(Cold stable)   & (62)   &  (2495)  &  (79) \\[2mm]
\hline
\end{tabular}
\end{center}
\label{tab:numbers}}
\end{table}

\subsubsection{Excess of luminous persistent LMBHB}
The number of systems for each model population is given in Table
\ref{tab:numbers}, which also gives the number of luminous, persistent
systems expected in each case according to the criterion
(\ref{eq:mcrit}).

The consistency of these numbers can be compared to observations,
which show no persistent LMBHB with short $P_{\rm orb}$. Possible
exceptions, i.e. systems showing variability that could be different
from what is called a transient behaviour, such as GX 339-4 or GRS
1915+105 have long orbital periods and are not addressed here. Model A
(standard MSW) predicts the highest number of persistent
sources. These 367 systems have an average luminosity $L=0.1
\dot{M}c^2 \approx 5 \cdot 10^{36}$\es. Given their location in the
Galaxy provided by the population synthesis code,  we estimate the
intervening absorption assuming that the gas is, roughly, distributed
as $n=1.8\exp(-d/3.5\mathrm{~kpc}-z/0.125\mathrm{~pc})
~\mathrm{cm}^{-3}$ with $d$ the distance to the Galactic centre, $z$
the height above Galactic plane and where the gas is distributed like
the thin disk population \citep{bahcall} with a density and
scale-height taken from \citet{zombek}. The average column density
along the line of sight from the Sun (at 8.5~kpc from the centre) 
to the systems is $\log N_H\approx 22.1$. 

 The accretion luminosity may be assumed to be emitted as that of
a 1~keV blackbody, corresponding to the `high/soft' X-ray spectral
state of LMBHB. Taking into account X-ray absorption, the number of
persistent systems expected above a threshold of 7 $\cdot 10^{-11}$
erg~s$^{-1}$~cm$^{-2}$ in the 2-10~keV band is 85. This threshold
corresponds to the completeness limit for the X-ray sources seen by
the \textit{ RXTE All Sky Monitor} (about 3 mCrab or 0.2 counts/s)
deduced from the $\log N - \log S$ distribution of extragalactic
sources by \citet{grimm02}. Setting the threshold ten times higher
still yields 13 bright, persistent LMBHB with \textit{ASM} fluxes
comparable to e.g. SMC X-1.

Assuming a power-law spectrum with photon index -2 (flat in $\nu
F_\nu$), corresponding to the `low/hard' X-ray spectral state of
LMBHB, one obtains almost the same number of detectable persistent
systems: a 1~keV blackbody peaks at 2.7~keV so the power-law emission
is not much different. Persistent LMBHB might escape detection if they
emit most of their energy below 1~keV. Indeed, KV UMa in the low/hard
state had blackbody-like emission peaking at $\approx 25$~eV, with a
spectral luminosity almost 10 times greater than in X-rays where it
displays a power law spectrum \citep{mcclintock2001}. Assuming that
only 10\% of the accretion luminosity is emitted in hard X-rays would
still imply 16 persistent systems above 3 mCrab in the \textit{RXTE
ASM} (1 above 30 mCrab).  Alternatively, the discs may be not
irradiated, in which case no systems would be stable; this, however,
would be surprising as both persistent NS LMXBSs and outbursting black
hole transients show clear signatures of irradiation. Note that with
parameter $\mathcal{C}=5\times 10^{-4}$ (Eq.~\ref{eq:tirr}), only a
very moderate 0.5\% of the accretion luminosity (efficiency of 10\%)
is reprocessed in the disc.

Despite obvious uncertainties, the large number of systems at high
mass-transfer rates in Model A makes it difficult to avoid the
presence of a significant number of observed persistent systems.
However, mass-transfer rates could be smaller than suggested by models
of binary evolution driven by AML through MSW and GWR. In such a case
Model B (GWR only) predicts no persistent systems at all while the
intermediate case represented by Model C (reduced MSW) yields 29
systems above 3~mCrab (or 5 above 30~mCrab) in the \textit{RXTE
ASM}. In addition mass-transfer rates predicted by model A are too
large compared to observations of transient systems. These points are
further discussed in the next section.

\subsubsection{Reduced Magnetic Braking}

While energy and angular momentum loss via gravitational wave
radiation is beyond doubt, the issue of angular momentum loss by
binaries via magnetically coupled stellar wind and its quantitative
description are much more arguable. The earliest formulation of Eq.
(\ref{eq:vz}) by \citet{vz81} and its variations are based on
extrapolation of stellar rotation braking law \citep{sku72} over
several orders of magnitude in stellar rotation rates, imply efficient
spin-orbit coupling for low-mass star and has been modified and/or
challenged by many authors \citep[][ and
others]{ham88,collier02,andron03,ivtaam03}; \citet{barkolb00} argued
that in order to explain the cataclysmic binary period distribution
(i.e. their evolution) a magnetic braking law according to which the
mass-transfer rate would, contrary to the ``standard" case, {\sl
decrease} with increasing orbital period might be necessary. It is
therefore fair to conclude that the so-called ``magnetic braking
mechanism" is so uncertain that it can be almost modified at will.

For this reason we computed a grid of evolutionary sequences for
binaries with low-mass donors and massive accretors under the
assumption that AML via MSW does not operate in semidetached bh+ms
systems. This choice was motivated by studies such as
\citet{andron03,ivtaam03} according to which the MSW operating in
close binaries is reduced by at least one order of magnitude with
respect to \citet{vz81} law; but mass-transfer values obtained in this
way should be rather treated as lower limits. Based on this set of
evolutionary tracks we derived the second model for the population of
semidetached binaries with black holes (model B). The model is
presented in the lower panel of Fig. \ref{fig:popul_las}.  In this
model the current Galactic population of semidetached bh+ms binaries
contains in total, $\approx 12000$ objects, about 5000 of which have
$q \geq 0.02$ (Table~\ref{tab:numbers}).

The absence of MSW AML results in lower \md\ in the systems with
similar initial masses of components and post-circularization
periods. As a result, in all systems of model B accretion discs are
unstable according to the criterion (\ref{eq:mcrit}).  Instability
limit is given by the lower bound of region filled by stable systems
in ``standard model''. In a model without magnetic braking in
semidetached systems typical mass-transfer rates $\dot{M}$ are by
factor up to 10 lower.

If MSW were to be absent from the very beginning, before the RLOF,
the results would not change much. The number of systems would be
reduced by a factor $\sim$ 2. However, the longest period systems
would be absent and this would make consistency with observations
worse.

The absence of stable bright low-mass sources argues in favour of an
AML reduced as compared to predictions of \citet{vz81} magnetic
braking model. It is worth noting that \citet{mnl99} also found that
black-hole SXTs evolution should be driven by AML without MSW. In
the view of both theoretical (the magnetic braking law, the value of
$\mathcal{C}$) and observational (the number of luminous, stable but
undetected systems) uncertainties it would not make much sense to
try to fix the value of parameter $f$ in Eq.~(\ref{eq:vz}) that would
bring the calculated population in exact agreement with
observations. An experiment with MSW angular momentum loss rate
reduced by factor 2 (Model C, see Table \ref{tab:numbers}) shows
that finding such an agreement will require a more substantial
reduction of the magnetic braking AML.

\subsection{Transient systems and cold, stable systems}

The prediction of our model can also be compared to mass-transfer
rates deduced from observations of SXT sources. Of course such
deductions are notoriously uncertain but, as we argue below, they
can provide interesting information especially when combined with
the knowledge that all observed short-period LMBHB are transient
X-ray systems.

\subsubsection{Reservoir discs}

The usual method for SXT consists
 of
obtaining the mass-transfer rate by
 dividing the mass accreted during outburst by the recurrence time
\citep[see e.g.][]{wp96}. Regretfully, among short-period SXT the
recurrence time is known only for A0620-00 (about 60 years) and 4U
1543-47 (about 10 years). For the other systems one can only obtain
upper limits by assuming that since only one outburst has been seen
the recurrence time is longer than 30 years. These estimates of \md,
derived from the data on outburst parameters from \citet{csl97} and
for KV UMa and V406 Vul based of observations presented in
\citet{chaty1118} and \citet{hynes1859} respectively, are given in the
penultimate column of Table \ref{tab:obs} and are shown in
Figs. \ref{fig:example}, \ref{fig:popul_las}, and \ref{fig:popul_mnl}.

\begin{table}
\caption[]{Orbital periods, mass-transfer rates based on recurrence
times and upper limits on the average mass-transfer rate for
black-hole candidate short-period binaries. See text for details. }
\begin{center}
\begin{tabular}{lrrc}
\hline\hline
Name  & $P_{orb}$,  & $\langle \md_{\rm recc} \rangle$,  & $\md_{\rm in}$,\\
      & day         &  \myr                                    & \myr \\
\hline
XTE J1118+480 (KV UMa)   & 0.171  &  $1.9 \times 10^{-12}$  & $1.5 \times 10^{-10}$\\
GRO J0422+32 (V518 Per)  & 0.212  &  $1.3 \times 10^{-11}$  & $2.3 \times 10^{-10}$\\
GRS 1009-45 (MM Vel)     & 0.285  &  $4.4 \times 10^{-11}$  & $3.8 \times 10^{-10}$\\
XTE J1650-500        & 0.320  &  $2.0 \times 10^{-11}$ & $4.8 \times 10^{-10}$ \\
A0620-00 (V616 Mon)      & 0.323  &  $3.3 \times 10^{-11}$  & $4.8 \times 10^{-10}$\\
GS 2000+25 (QZ Vul)       & 0.345  &  $2.0 \times 10^{-10}$  & $5.4 \times 10^{-10}$\\
XTE J1859+226 (V406 Vul) & 0.383  &  $4.1 \times 10^{-10}$  & $6.4 \times 10^{-10}$\\
GRS 1124-68 (GU Mus)      & 0.433  &  $3.4 \times 10^{-10}$  & $8.0 \times 10^{-10}$\\
H 1705-25 (V2107 Oph)     & 0.521  &  $5.5 \times 10^{-11}$  & $1.1 \times 10^{-9} $\\
4U 1543-47  (IL Lup)      & 1.125  &  $4.2 \times 10^{-10}$  & $4.3 \times 10^{-9} $\\
\hline
\end{tabular}
\end{center}
\label{tab:obs}
\end{table}

Clearly even pure GWR cannot explain such low mass-transfer rates if
one assumes that they represent {\sl secular} values. No LMBHB with
a period $\apgt 4$ hr can have a secular mass-transfer rate as low
as $\md \sim 10^{-12}$\,\myr; it would take post-minimum period
systems several Hubble times to reach these periods and
corresponding mass-transfer rates. The only known type of systems
where such \md\ may be expected, are binaries with initially
degenerate dwarf donors, but even then orbital periods should be
$\lesssim 2$ hr \citep{del_arbitr05}. If the extremely low
mass-transfer rate deduced for short period SXTs were due to
downward fluctuations from   an intrinsically high secular
 mass-transfer rate, 
one would still have to explain why the bright counterparts are not
observed (Section \ref{sec:dimstable1}). Furthermore, some systems
would then necessarily have higher rates than their secular value
(by how much depends on the duty cycle of the purposed low states),
hence there should be even more persistent systems.

For longer periods, Fig. \ref{fig:example} shows, \md\ of systems
with $P_{\rm orb} \approx 8 - 10$\, hr may be consistent with
evolution of systems where donors overflow Roche lobes extremely
close to TAMS, but then one needs to assume some very special
initial distribution of binaries over orbital separations which will
finally, after several evolutionary stages with mass transfer and
mass and momentum loss from the system and supernova explosion,
result in concentration of zero-age black-hole and
main-sequence-star systems just at the desired very narrow range of
separations.

The longest period SXT of  the ``short-period'' group, 4U1543-47 is, in
principle, consistent with  the evolution of systems with donors
overflowing Roche-lobes after formation of He-cores. But again,
these binaries evolve through  the Hertzsprung gap extremely fast, and
 the probability of observing a descendant of such systems is very low, see
discussion of intermediate-mass ``Hertzsprung gap'' black-hole
binaries in \citet{kolb_soft98}.

\subsubsection{Truncated discs \label{sec:mnlpar}}

It is, however, plausible that the mass-transfer rates in
short-period SXTs are much higher than those derived above. The
method used to obtain the low values mentioned above treats the disc
as a reservoir which during quiescence is filled up to a critical
value at which the outburst is triggered and the disc emptied. The
implicit assumption in this picture is that the reservoir is not
leaky, i.e. that the accretion rate at the disc's inner edge is much
smaller than the mass-transfer rate, as indeed predicted by the DIM
when the disc extends to the innermost stable circular orbit.
However, there are many reasons to think that quiescent \citep[and
not only quiescent, see e.g.][]{done_gier06} SXT discs are truncated
and therefore ``leaky" because of the inner truncation required to
explain observed quiescent luminosities \citep[e.g.][]{l96,letal96}.
Also, \citet{dubhamlas01} showed that the disc instability model can
reproduce the observed X-ray light-curves only if discs in SXTs are
truncated (and irradiated). The truncated disc ``paradigm" is now
commonly used in describing the observed timing and spectral
properties of X-ray LMBHB \citep[][ and references therein]{mclrbh}.

For systems with {\sl non-stationary} quiescent accretion discs the
mass-transfer rate can be estimated as \citep{lasota01}
\begin{equation}
{\dot M_{\rm tr}\approx \frac{\epsilon M_{\rm D, max}}{t_{\rm
recc}}+ \dot M_{\rm in}}; \label{eq:mdot}
\end{equation}
where
\begin{equation}
M_{\rm D, max}=2.7 \times 10^{21} \alpha^{-0.83} \left( \frac{M_1}
{\rm M_\odot} \right)^{-0.38} \left( \frac{R_d}{10^{10} \; \rm cm}
\right)^{3.14} \ {\rm g}, \label{eq:diskmass}
\end{equation}
is the maximum quiescent-disc mass and $\epsilon= \Delta M_{\rm
D}/M_{\rm D, max}$ the fraction of the disc's mass lost during
outburst; $\alpha$ is the kinematic viscosity parameter, $R_d$ is
the disc's outer radius and $\dot M_{\rm in}$ is the accretion rate
at the disc's inner edge. The usual, non-leaky disc estimates,
neglect $\dot M_{\rm in}$. For a stable equilibrium disc $t_{\rm
recc}= \infty $ and $\dot M_{\rm tr}= \dot M_{\rm in}$.

However, in SXTs $\dot M_{\rm in}$ is in fact the dominant term as
argued by \citet{mnl99}. They applied models in which the inner part
of the accretion flow was represented by an ADAF
\citep[][]{ny94,abrameteal95} to several quiescent SXT systems.
Fitting the model to the observed spectra provided estimates of
$\dot M_{\rm in}= \dot M_{\rm ADAF}$ which are higher than the
``refilling" term $\epsilon M_{\rm D, max}/t_{\rm recc}$, for which
except for A0620-00 (and the long period SXT system V404 Cyg) only
upper limits could be obtained.

\citet{lasota00} suggested a general estimate of $\dot M_{\rm in}$
in quiescent SXTs. It is based on the fact that in quiescence
according to the DIM the disc accretion rate has to satisfy the
inequality:
\begin{equation}
\md(r) < \md^-_{\rm crit} \approx 6.4 \times 10^{-11} \left(
\frac{M_1} {\rm M_\odot} \right)^{-0.88} \left(\frac{r}{10^{10} \rm
~cm}\right)^{2.65} \myr \label{eq:mcritcold}
\end{equation}
\citep[][note that this criterion which requires the whole
\textsl{quiescent} disc to be in a cold thermal equilibrium state is
different from the instability condition of Eq.
\ref{eq:mcrit}]{dubus_phd,lasota01}. Therefore, the mass accretion
rate at the truncation radius ($\dot M_{\rm in}$) must be smaller
than $\md^-_{\rm crit}(r_{\rm in})$. Inside the truncation radius
the mass accretion rate $\dot M$ is constant ($\dot M=\dot M_{\rm
in}$), the accretion timescale in the ADAF (or similar type of flow)
being very short. Parameterizing the truncation radius as a fraction
of the circularization radius \citep{fkr02} $r_{\rm in}=f_t r_{\rm
circ}$ \citep[where $f_t < 0.48$, see][]{mnl99} and assuming a mass
ratio $q \approx 0.1$ one obtains
\begin{equation}
\dot M_{\rm in} \lta  10^{-8} f_t^{2.65} P_{\rm day}^{1.77} {\ms \
\rm y^{-1}}, \label{eq:mdot_adaf}
\end{equation}
which because of the assumed value of $q$ is independent of the
primary's mass. Taking $f_t \approx 0.48$ one obtains
\begin{equation}
\dot M_{\rm in} \lta 3.6 \times 10^{-9} P_{\rm day}^{1.77} {\ms \
\rm y^{-1}}. \label{eq:mdot_adaf2}
\end{equation}
which, considering the uncertainties, corresponds reasonably well to
the observed $L_X(P_{\rm orb})$ relation for quiescent SXTs
\citep{ham03}, or at least can be considered as an upper limit.

Using Eq. (\ref{eq:mdot_adaf2}) one obtains upper limits to
mass-transfer rates given in the last column of Table \ref{tab:obs}.
They are consistent with the model B and one can see that a small
amount of additional AML could provide an even better agreement. One
can therefore conclude that a substantially reduced MSW AML would be
consistent with both the stability properties and mass-transfer
rates of SXTs.

\subsubsection{Number of transients}
Estimates of the number of SXTs in our Galaxy based on observations
range from several hundreds to a couple thousand
\citep{vdH1992,csl97,romani1998}. The number of SXTs predicted by
the models is 3000 (Model A) to 5000 (Model B), on the upper end of
the estimates based on observations. Assuming that their outbursts
peak on average at 0.1~$L_{\rm edd}$ in X-rays, all of these
transients should be seen in outburst by the \textit{RXTE ASM},
regardless of their location in the Galaxy. The current discovery
rate seems to be $\approx$ 2--5 per year. Matching this rate would
imply recurrence  times of several hundred years. \citet{csl97}
find that the average peak luminosity of transients is $\log
L\approx 37.7\pm0.8$. The peak flux could be overestimated: with a
peak at $10^{37}$ \es\ in the \textit{RXTE ASM} bandpass, only
$\approx$ half of the systems are above a 30~mCrab threshold
(appropriate for transients) after absorption, yielding recurrence
rates closer to $\approx 100$~years. Given the relatively high
number of transients expected, the population synthesis taken at
face value would suggest that many transients are sub-Eddington at
maximum and/or that recurrence times are $\ga 100$~years for most
sources.

\begin{figure}
\center
\includegraphics[angle=-90,width=\columnwidth]{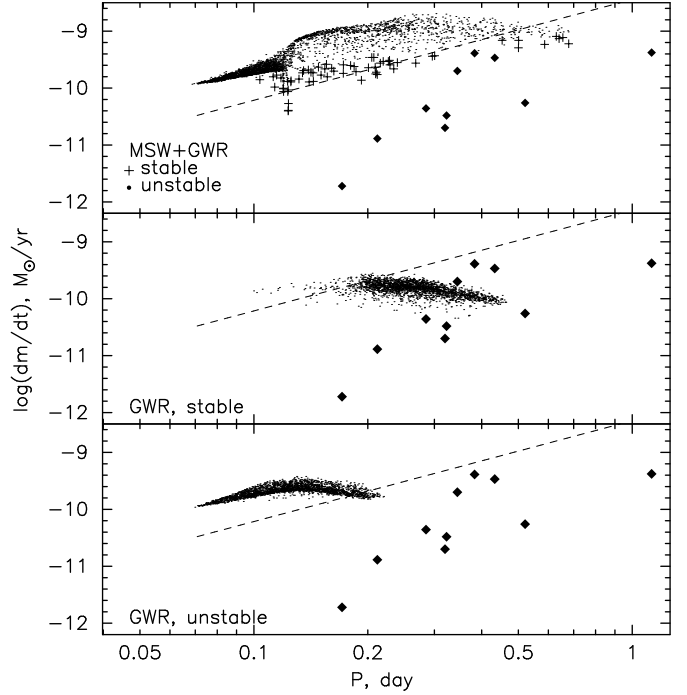}
\caption[]{Mass transfer rates vs. orbital period for the model
population of semidetached low-mass binaries with assumed black hole
accretors. Upper panel -- ``standard'' evolution model with disrupted
magnetic braking braking (Model A).  Pluses mark stable according to
Menou et al.  stability criterion for cold discs systems, while dots
-- unstable ones.  Middle panel shows position of systems with discs
stable acording to Menou et al. criterion in the model in which MSW
does not operate in semidetached systems with black hole accretors
(model B).  Lower panel -- systems with unstable discs in model B.
Diamonds are mass-transfer rates estimated on the base of assumed
recurrence periods, like in Fig. \ref{fig:popul_las}.}
\label{fig:popul_mnl}
\end{figure}

\subsubsection{A population of faint LMBHB?}

A model by \citet{mnl99} (MNL) may be the answer to this apparent
difficulty. Following a suggestion by \citet{letal96} they pointed out
that if the truncation radii were slightly larger than estimated to
fit observations of quiescent SXTs the discs would be globally
stable. Indeed, stationary (\md = const.) discs truncated at
\begin{equation}
R_{\rm in} > R_{\rm crit} = 1.69 \times 10^9 \left(\frac{\dot
M}{10^{-10} \msol}\right)^{0.375} \left( \frac{M_1} {\rm M_\odot}
\right)^{0.33} \label{eq:rcrit}
\end{equation}
are cold and globally stable since their accretion rates are lower
than the critical value given by Eq. (\ref{eq:mcritcold})
corresponding to cold stable equilibria. Hence, some of the systems
classified by us as transients would instead be cold, stably accreting
sources. \citet{mnl99} suggested that a population of such persistent
LMBHB could be present in the Galaxy. Despite their relatively large
mass-transfer rates ($\dot M \sim 10^{-10}\ {\ms \ {\rm yr}^{-1}}$)
these systems would be very faint and difficult to detect so it is
interesting to establish if their putative existence is compatible
with population models.

Figure \ref{fig:popul_mnl} shows the distribution of stars in two
population models with respect to stability criterion suggested by
MNL. Here, we assume the truncation occurs at $R_{in}= 0.48 R_{\rm
circ}$, the radius at which overflowing material from the donor star
merges with the accretion disc \citep{mnl99}. MNL-stable systems have
$R_{in} > R_{\rm crit}$. The break-down over MNL-stable and unstable
(transient) systems is given in Table \ref{tab:numbers}. In the
``standard'' case with MSW AML (Model A) very few discs that were
unstable vs.  DIM-criterion, become stable vs. MNL-criterion.  On the
other hand, about a half of DIM unstable systems become stable vs. the
MNL criterion in the model without MSW after RLOF (Model B). This is
because mass transfer rates in the GWR-only scenario are much lower
than in the MSW+GWR scenario, thereby enabling the criterion for a
cold disc to be met more easily. The number of transients for both
models stays at $\approx$ 2000-3000.  On the other hand, truncating
the disc at R$_{\rm circ}$ would make all transients ``MNL-stable''
for both models. It is therefore a distinct possibility that there are
very few ``classical'' SXTs in the DIM sense (i.e. regularly
outbursting) and that the LMBHB population divides up into hot/stable
and cold/stable systems, the latter erupting from time to time due to
weather variability in truncation physics or mass transfer rate. Even
in this extreme case, the predicted population size is too small to
have cold/stable systems contribute significantly to the large number
of faint ($L_X\sim 10^{31}$~erg~s$^{-1}$) hard X-ray sources seen in
deep \textit{Chandra} exposures toward the Galactic Center
\citep{muno03,belczynski04}.

\subsection{Systems with very low mass ratios}
\label{sec:low_q}

Mass transfer in systems with very low mass ratios is uncertain and
they have been left out of the discussion until now. For values of
$q\la 0.02$ the circularization radius becomes greater than the
estimates of the outer radius given by \citet{Paczynski77} and
\citet{pappringle77}. Matter flowing in from the companion
circularises onto unstable orbits. At $q\approx 0.02$, matter is
added at $R_{\rm circ}$ onto orbits that can become eccentric due to
the 3:1 resonance. At $q\approx 0.005$ the circularization radius
approaches the 2:1 Lindblad resonance. This can efficiently prevent
mass being transferred onto the compact object (by analogy with the
gaps created by planets in protoplanetary discs or the inner edges
of circumbinary discs) unless the timescale of mass transfer is such
that the gap can be filled before it is cleared (unlikely, as the
condition for gap formation is roughly $M_2/M_\odot \ga t_{\rm
dyn}/t_{\rm vis}$,  where $t_{\rm dyn}$ and $t_{\rm vis}$ are
respectively the dynamical and viscous timescales). Subsequent
evolution is probably set by the balance between GW radiation
(dominating at low $q$) which drives the system inward and the
torque exerted by the disc which drives the system outward.
Regardless of this exact balance, since the outer radius of the
accretion disc is unlikely to be greater than $R_{\rm circ}$, most
of these systems (if accreting) are likely to be MNL stable
according to the discussion in the preceding section.

\section{Discussion and Conclusions}
\label{sec:disc}

We have modeled populations of short-period semi-detached close
binary systems containing black holes and low-mass stars under
various assumptions about the AML mechanism. In all cases the
obtained population can be divided into two subsets. One, contains
short-period ($\lta 0.1$ day), very low mass-ratio ($q < 0.02$)
systems; the other contains systems whose orbital parameters
correspond to those determined for black hole binary transients.
Mass transfer in short-period systems may be interrupted by
resonances within the primary's Roche lobe and their subsequent
evolution and the observational consequences are uncertain. For the
systems with mass ratios $q>0.02$, the disk instability model
provides a framework to investigate the observational consequences
in terms of bright, persistent systems and transients.

Among systems with the range of periods encountered in SXTs (let us
note that the shortest observed period of a LMBHB is 4.1 hours but
nothing in our model prevents such systems to have periods $\sim 2$
hours) the number of transient systems depends on the assumed AML
mechanism. The standard Verbunt \& Zwaan (1981) formula implies the
presence of luminous stable systems that are not observed. On the
other hand, AML driven solely by GWR yields only transient
systems. This suggests that in short-period LMBHB the strength of the
AML MSW mechanism is significantly reduced. Such a reduction is also
required by the values of mass-transfer rates deduced from
observations. Our population model shows that the required
mass-transfer rates cannot be due to the evolved nature of the
mass-donor.

The number of transients in the models is in the few thousand range,
at the upper end of estimates based on observations. Better
agreement would imply transients having lower peak fluxes and/or
long recurrence times. Another possibility is that the number of
transients is reduced compared to these estimates. A significant
fraction of transients could be cold and stable if the thin
accretion disc is truncated in quiescence. The number of such
systems is highest with GWR-only angular momentum losses (lower mass
transfer rates): with $R_{\rm trunc}=0.48 R_{\rm circ}$, roughly
half of transient SXTs then become cold and stable. With MSW+GWR,
the number of cold, stable systems is very small. This fraction can
be increased to include all systems if the disc is truncated up to
$R_{\rm circ}$. If this were the case, transients would be flukes
(not triggered by the DIM but by external factors) and the regularly
outbursting SXTs envisioned by the DIM would be an exception rather
than the rule as suggested by \citet{letal96}.

Recently, \citet{podsetal03} and \citet{justham_bh06} suggested that
low-mass companion stars are not likely to provide sufficient
gravitational potential to unbind during common-envelope phase the
envelope of black-hole's massive progenitor. In our formulation,
this is equivalent to a low value of the product of common-envelope
efficiency and structure parameter $\alpha\lambda$. A trial
computation with $\alpha\lambda=0.1$ resulted in a population with
only $\simeq 3400$\  short-period binaries containing black holes
with low-mass companions. Only the most massive binaries avoided
merger due to the fact that they lose more mass on the main sequence
and thus have rather low-mass envelopes and survive the
common-envelope phase. However, all black holes produced in this
case have mass exceeding $\simeq 14$\,\ms, substantially higher than
what is observed in short-period LMBHB. Post-circularization period
-- secondary mass relations for these binaries are similar to these
shown in Fig. \ref{fig:p0m20}. Thus, if such systems were formed, we
expect evolution similar to described in  this  paper, but without
formation of LMBHB resembling observed ones. However, it is
difficult to reach firm conclusions on this issue as both the
details of the common-envelope phase as well as the formation of
black holes are are very uncertain. So low-mass stars may still be
able to remove the envelope or if indeed only the most massive
binaries survive, their black hole masses may be lower and be
consistent with the observed masses

We conclude that modeling the population of LMBHB, using 
standard models of stellar evolution and accretion physics, leads to
prediction of (1) a significant population of systems with $q<0.02$
and $P_{\rm orb}\le$ 2 hours; (2) a significant number of persistent
LMBHB unless MSW angular momentum losses are reduced; (3) a high
number of transients unless the peak X-ray fluxes are low and/or the
recurrence time high; (4)  a possibility of having a large
fraction, if not all LMBHB systems, actually being (secularly) cold
and stable.

\section*{Acknowledgments}

We thank Peter Eggleton for providing the copy of his evolutionary
code and  Samuel Boissier for providing the table of star formation
rates and helpful discussion about the Galactic model. GN is
supported by PPARC and NWO. LRY is supported by NWO, NOVA, RFBR, and
Russian Academy of Sciences Basic Research Program ``Origin and
Evolution of Stars and Galaxies''. LRY acknowledges warm hospitality
and support from Institut d'Astrophysique de Paris, Universit\'e
Pierre et Marie Curie, and Astronomical Institute ``Anton
Pannekoek'', where a part of this study was carried out. GD, JPL and
LRY were supported by grants from the Centre National d'\'Etudes
Spatiales and the GDR PCHE of the CNRS. SPZ acknowledges support
from KNAW and LKBF.

\bibliographystyle{aa}

\appendix

\section{Model for the star formation history and stellar density distribution}\label{appendix_model}

The probability of a star being born at a radius $R$ or smaller
  depends on the integrated SFR, i.e.,
\begin{equation}
P(R, t) = \frac{\int_0^R \mbox{SFR}(R', t) \, 2 \, \pi \, R' \, {\rm d} R'}{\int_0^{R_{\rm max}} \mbox{SFR}(R',
  t) \, 2 \, \pi \, R' \, {\rm d} R'},
\end{equation}
where $R_{\rm max}$ is the maximum extent of the Galactic disc (19 kpc). We
assume that the stars do not migrate radially in time. The age of the
Galaxy in this model is 13.5 Gyr. Since the \citet{bp99} model gives the
SFR projected onto the plane of the Galaxy, we assume a $z$-distribution in the Galactic disc
\begin{equation}\label{eq:Pz}
P(z) \propto \mbox{sech}(z/z_h)^2,
\end{equation}
where $z_h$ = 200 pc, neglecting dependence on age and
mass. We have added a bulge to the model of Galactic distribution of stars, by
doubling the SFR in the inner 3 kpc of the Galaxy compared to
\citet{bp99}. The total mass in this region at the age of 13.5 Gyr is then
2.6 10$^{10}$ \ms, consistent with
kinematic and micro-lensing results \citep[e. g.,][]{kzs02}. We
distribute the stars in the bulge according to
\begin{equation}
\rho_{\rm bulge}  = \exp(-(r/0.5 \mbox{ kpc})^2),
\end{equation}
where $r = \sqrt{x^2 + y^2 + z^2}$. The time-dependence of SFR under given assumptions and comparison to often used exponential SFR may be found in Fig. 2 of \citet{nyp04}.

\bibliography{4984}

\end{document}